\documentclass[11pt]{article}
\usepackage{moriond,epsfig,subfigure,floatflt}

\def\mco{\multicolumn}

\def\be{\begin{equation}}
\def\ee{\end{equation}}
\def\bea{\begin{eqnarray}}
\def\eea{\end{eqnarray}}

\begin{document}
\vspace*{4cm}
\title{LEPTON STUDIES WITH ATLAS AND CMS EARLY DATA}

\author{ C. ADAM-BOURDARIOS }
\address{Laboratoire de l'Acc\'el\'erateur Lin\'eaire, Orsay, France }

\author{ ON BEHALF OF THE ATLAS AND CMS COLLABORATIONS \footnote{Talk given at the Rencontres de Moriond, QCD, March 14-21 2009, ATL-PHYS-PROC-2009-049 } }
\address{ }

\maketitle\abstracts{
Leptons will play a key role in the first measurements performed by the LHC experiments. The ATLAS and CMS detectors are now built, and early estimates of their performance have been revisited. The expected first physics signals are described: they will shed light on the level of detector understanding and allow a control of the systematics uncertainties, opening the road to new and precision physics.
}

The last two years have seen the end of the ATLAS and CMS detector assembly. This allowed a survey of the detector elements placement accuracy, and a precise cross-check of the amount and position of the dead material. All of this was gradually integrated in the detector simulation, which is now believed to be realistic. The Geant 4 simulation and the calibration procedure have been extensively validated by a set of Test-Beams. Cosmics data, taken in summer 2008, are being used to study and improve the detector alignment~\cite{commissionning}. Last but not least, the reconstruction algorithms have now stabilised, in  particular for lepton identification.

For all these reasons, early estimates made over the years were revisited. Both collaborations have published detailed reports about the expected performance, to be used as references and guidelines in the coming years~\cite{ATLAS-paper,CSC-ATLAS,TDR-CMS}. They are both now focusing on the short term physics program, accessible with integrated luminosities of about 100 pb$^{-1}$.

\section{Detector Performance for lepton identification}

\subsection{Electrons }\label{subsec:electron}

The first challenge for electron reconstruction is the resolution and linearity of the energy measurement. After years of very detailed calibration studies, the analysis of Test-Beam data has proven that the design goals were met, with a local constant term smaller than 0.5~$\%$.  
The second challenge is the jet rejection: QCD cross sections are such that one expects an electron to jet ratio of $10^{-5}$. Both experiments have developed cut-based algorithms, meant to be robust enough to be used at the startup. The efficiencies and rejections obtained by ATLAS are shown in Table~\ref{effic-elec} ~\cite{CSC-ATLAS}: eventually, multivariate techniques will allow to increase the efficiency by 4 to 8~$\%$, and the rejection by 40 to 60~$\%$.

\begin{table}[]
\label{effic-elec}
\caption{Electron identification efficiency and jet rejection of the ATLAS cut-based algorithm, averaged over all $E_T > 17$ GeV. The jet rejection factor needed ( 10$^5$ ) can be achieved via two sets of cuts. }
\begin{center}
\begin{tabular}{|c|c|c|l|}
\hline
Cuts & \mco{2}{|c|}{Efficiency ($\%$)} & Jet rejection \\
\hline
& Isolated electrons & Electrons in jets & \\
\hline
Loose & 87.96 $\pm$ 0.07 & 50.8 $\pm$ 0.5 & 567 $\pm$ 1 \\
Medium & 77.29 $\pm$ 0.06 & 30.7 $\pm$ 0.5 & 2184 $\pm$ 13 \\
Tight(TRT.) & 61.66 $\pm$ 0.07 & 22.5 $\pm$ 0.4 & (8.9 $\pm$ 0.3) $10 ^4$ \\
Tight (isol.) & 64.22 $\pm$ 0.07 & 17.3 $\pm$ 0.4 & (9.8 $\pm$ 0.4)  $10 ^4$ \\
\hline
\end{tabular}
\end{center}
\end{table}

\subsection{Muons }\label{subsec:muon}

\begin{figure}[]
\caption {ATLAS muon identification performance in physics events, as a function of the pseudo-rapidity $\eta$. }
\label{fig:effic-muon}
\begin{center}
\subfigure[muon reconstruction efficiency]{ \includegraphics[width=45mm,height=35mm]{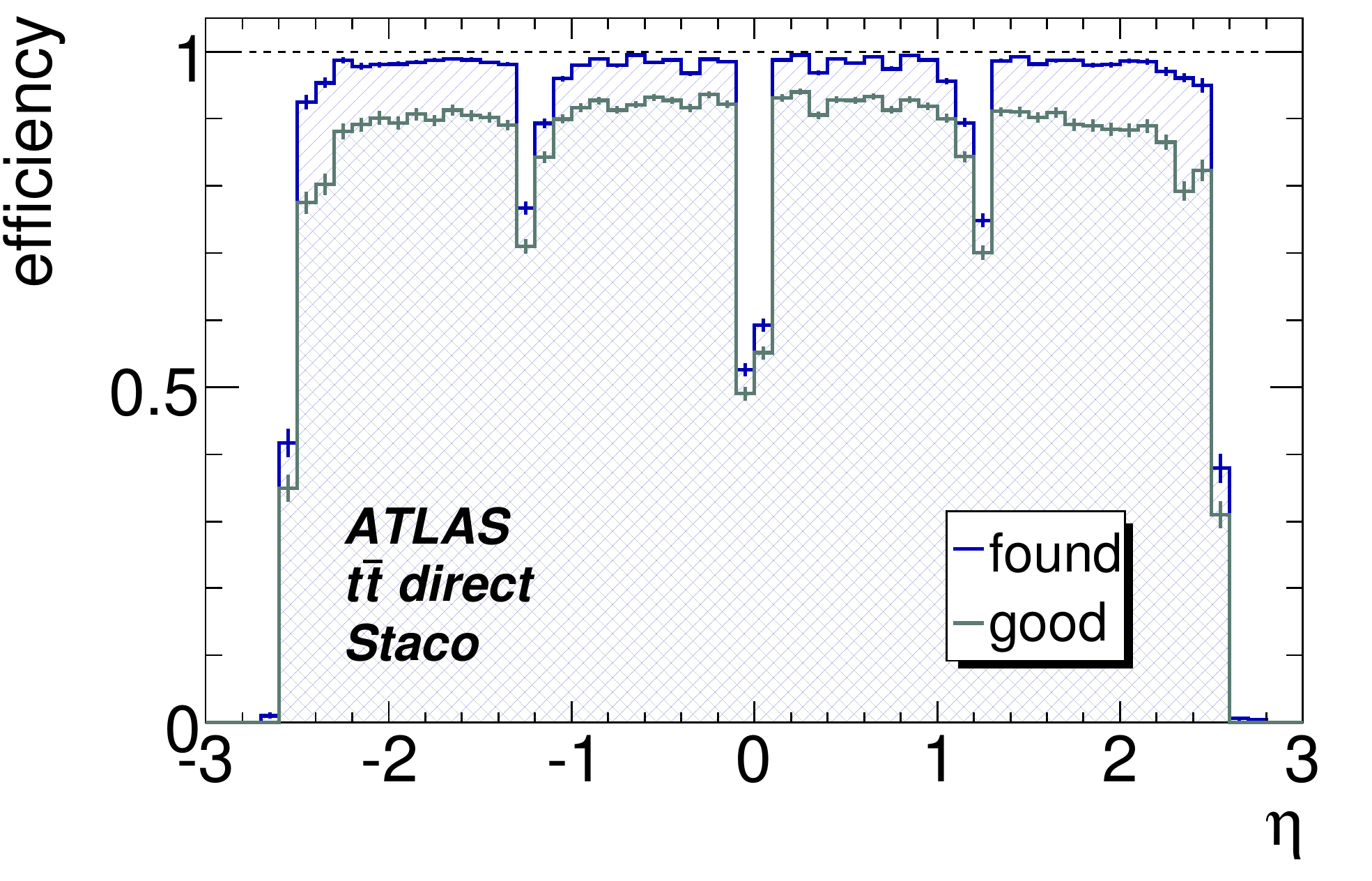} }
\subfigure[number of fake muons per event]{ \includegraphics[width=45mm,height=35mm]{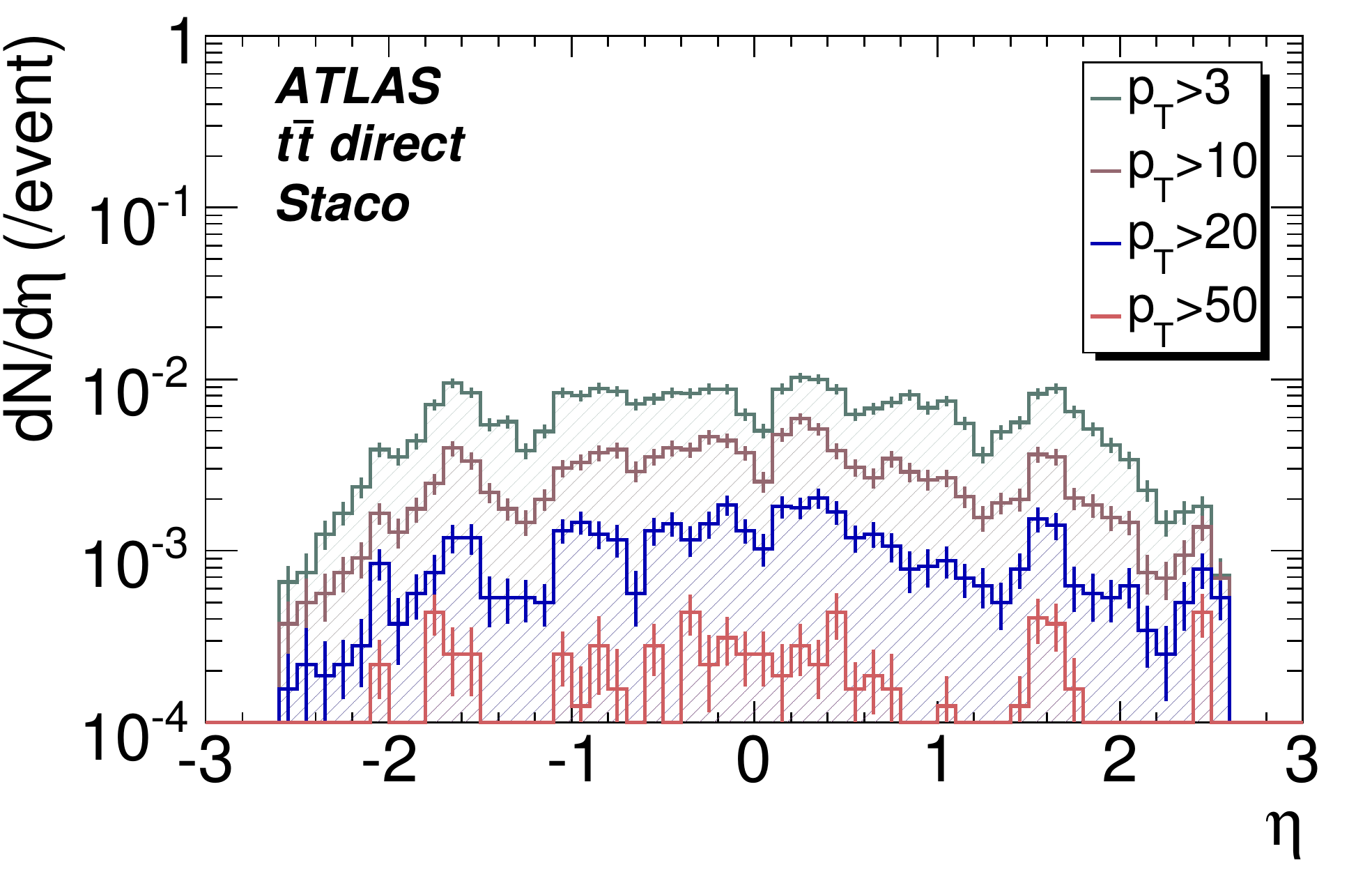} }
\end{center}
\end{figure}

Both experiments use a combination of the Tracking and Muon detectors to identify and measure their momentum between 4 GeV and 1 TeV. 
Figure~\ref{fig:effic-muon} shows the ATLAS muon identification efficiency and fake rejection~\cite{CSC-ATLAS}: for $P_T> 20$ GeV, one expects less that $2.10^{-3}$ fake muons per event.

The J/$\psi$ will be one of the first signals reconstructed at the LHC startup. The CMS collaboration expects, with only 3 pb$^{-1}$, about $7.10^4$ $J/\psi \rightarrow \mu^+ \mu^-$ reconstructed decays, with transverse momenta between 5 and 40 GeV. Making use of the large B lifetime, the promptly produced J/$\psi$ will be separated from those issued from B decays, and the cross section measured with a systematic uncertainty of 15~$\%$. But the reconstructed dimuon invariant mass will also be a probe of the alignment and muon reconstruction quality: it is expected to be 34.2 MeV at the startup, compared to 29.5 MeV with an ideal alignment~\cite{CMS-notes}.

\subsection{Taus }\label{subsec:taus}

$\tau$ reconstruction is tricky and relies ( not for the first data but soon after ) on multivariate techniques. ATLAS has followed two approaches: a calorimeter based and a track based algorithm. Typical efficiencies are 30~$\%$ for a jet rejection of about $10 ^3$. With 100 pb$^{-1}$ one expects clear signals for Z and W in $\tau$ decays. These will eventually be used to set the transverse missing energy scale to a few $\%$ ~\cite{CSC-ATLAS}.

\section{Z and W leptonic decays reconstruction }

\begin{figure}[]
\caption { Reconstructed Z mass in leptonic decays }
\label{fig:Zreco}
\begin{center}
\subfigure[Electron channel]{ \includegraphics[width=45mm,height=40mm]{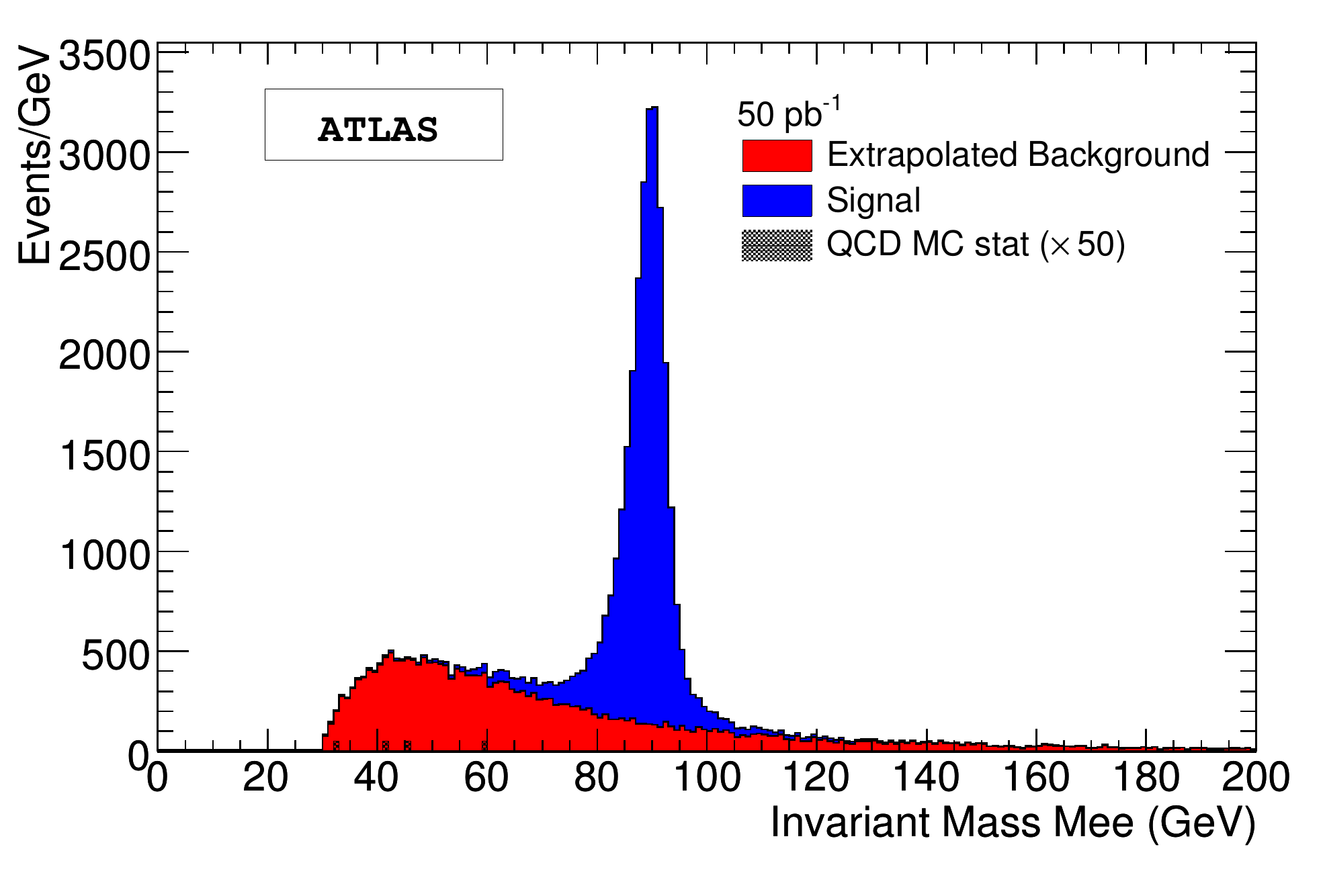} }
\subfigure[Muon channel]{ \includegraphics[width=50mm,height=45mm]{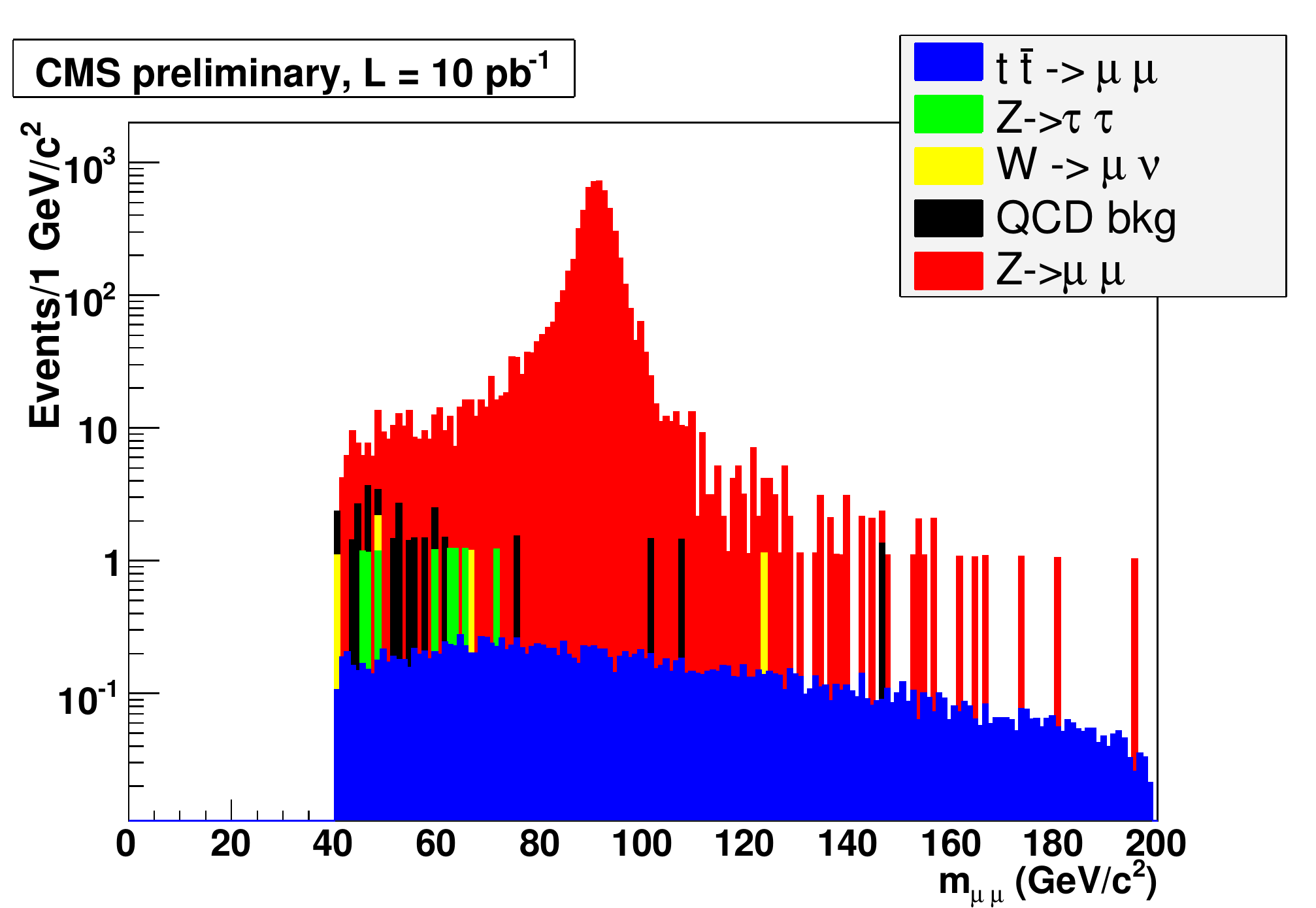} }
\subfigure[Tau channel]{ \includegraphics[width=45mm,height=38mm]{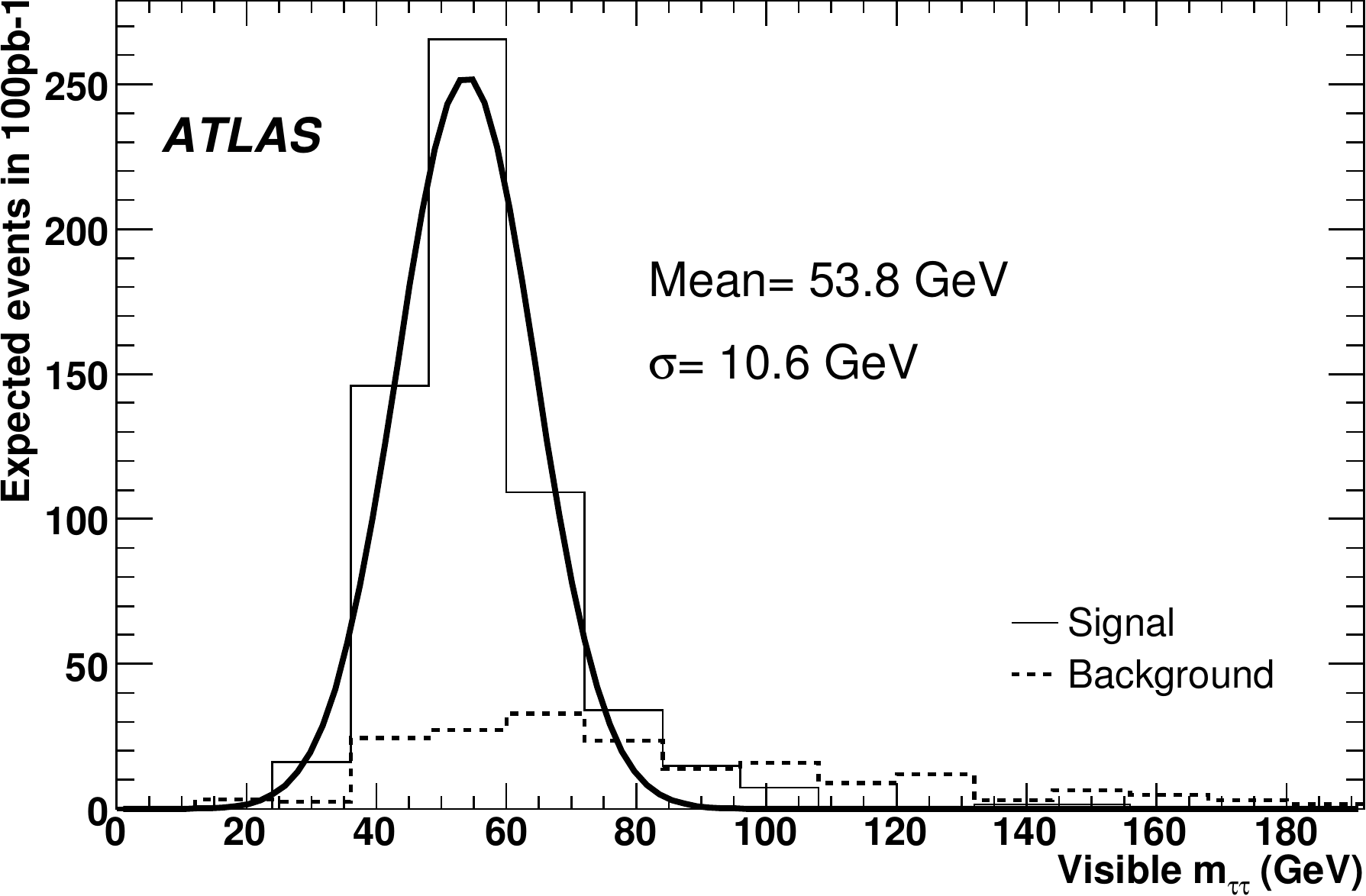} }
\end{center}
\end{figure}

The W and Z production cross sections are proportional to the beam energy and ATLAS expects, at 14 TeV and for 50 pb$^{-1}$, 2.48 $10^4$ $Z \rightarrow e^+ e^-$ and 2.56 $10^4$ $Z \rightarrow \mu^+ \mu^- $ reconstructed decays. Figure~\ref{fig:Zreco} shows the reconstructed Z mass in the three leptonic decay channels. These Z decays are often called candles, because they will initially be used to shed light on our detector understanding, via the tag and probe method: a tight identification is requested for one of the leptons; the second lepton can then be used to measure the efficiency of any identification cut. With 100 pb$^{-1}$, such studies will allow to reach an accuracy of about 2~$\%$.

Z decays will also be used to spot large scale inhomogeneities that the local calibration would not have seen, differences in the energy scale of different detector pieces, and residual misalignment. With 100 pb$^{-1}$, ATLAS expects to have enough reconstructed Z decays to check the muon momentum resolution and the electron energy scale at the per cent level. 

W leptonic decays are selected by the presence of an identified isolated lepton, of high transverse momentum ( typically above 20 GeV ), and some missing energy due to the escaping neutrino ( typically above 25 GeV ). Assuming NLO cross sections, with 10 pb$^{-1}$, CMS expects about 28 $10^3$ $W \rightarrow e \nu $ and 64 $10^3$ $W \rightarrow \mu \nu$ events, the dominant background being QCD events. 

\begin{figure}[]
\caption { Transverse mass ($M_T$) distribution }
\label{fig:transverse-mass}
\begin{center}
\subfigure[$W \rightarrow \mu \nu$ channel]{ \includegraphics[width=45mm,height=40mm]{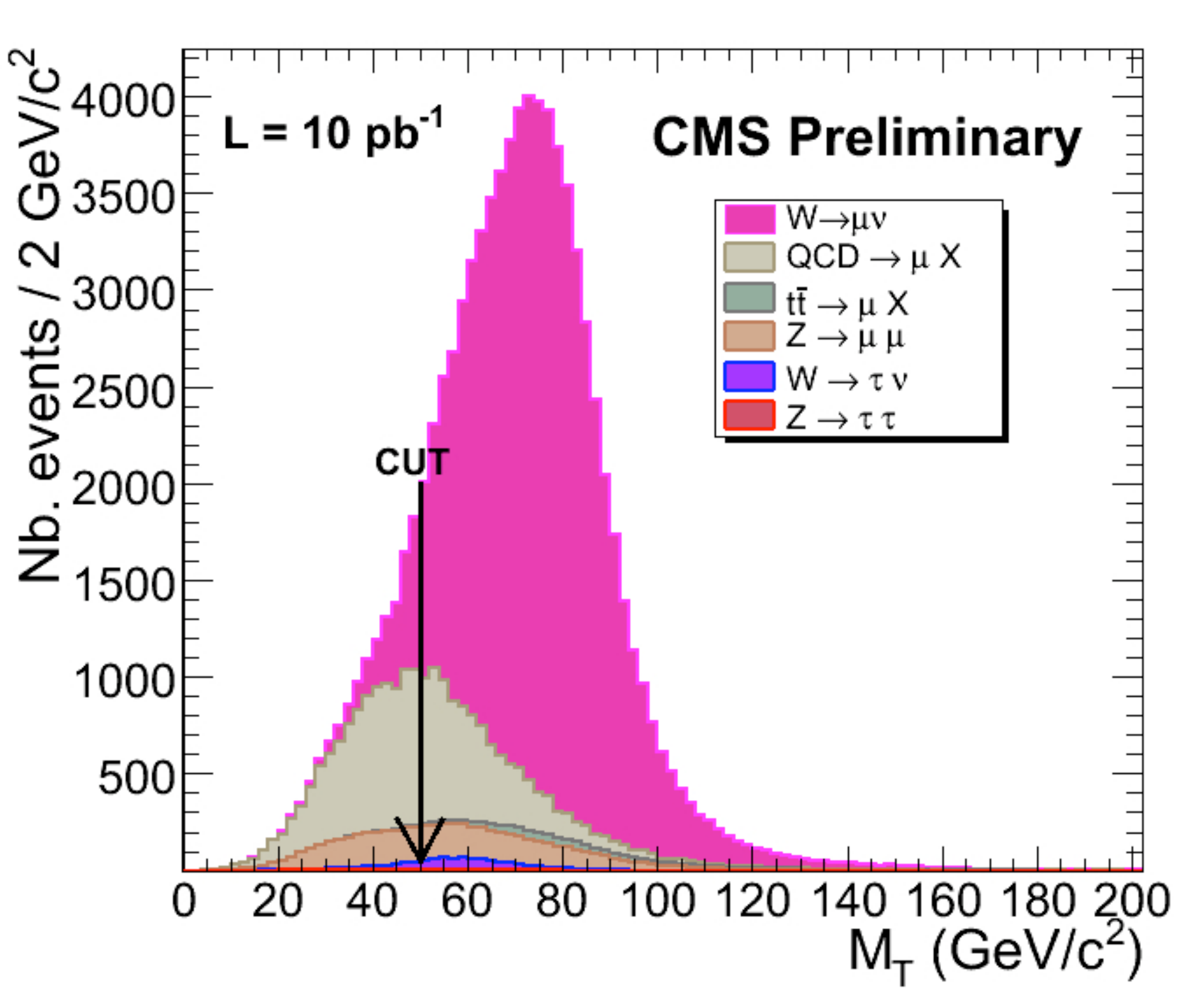} }
\subfigure[Comparison between real and fake W decays, produced using $Z \rightarrow \mu \mu$ events ]{ \includegraphics[width=45mm,height=40mm]{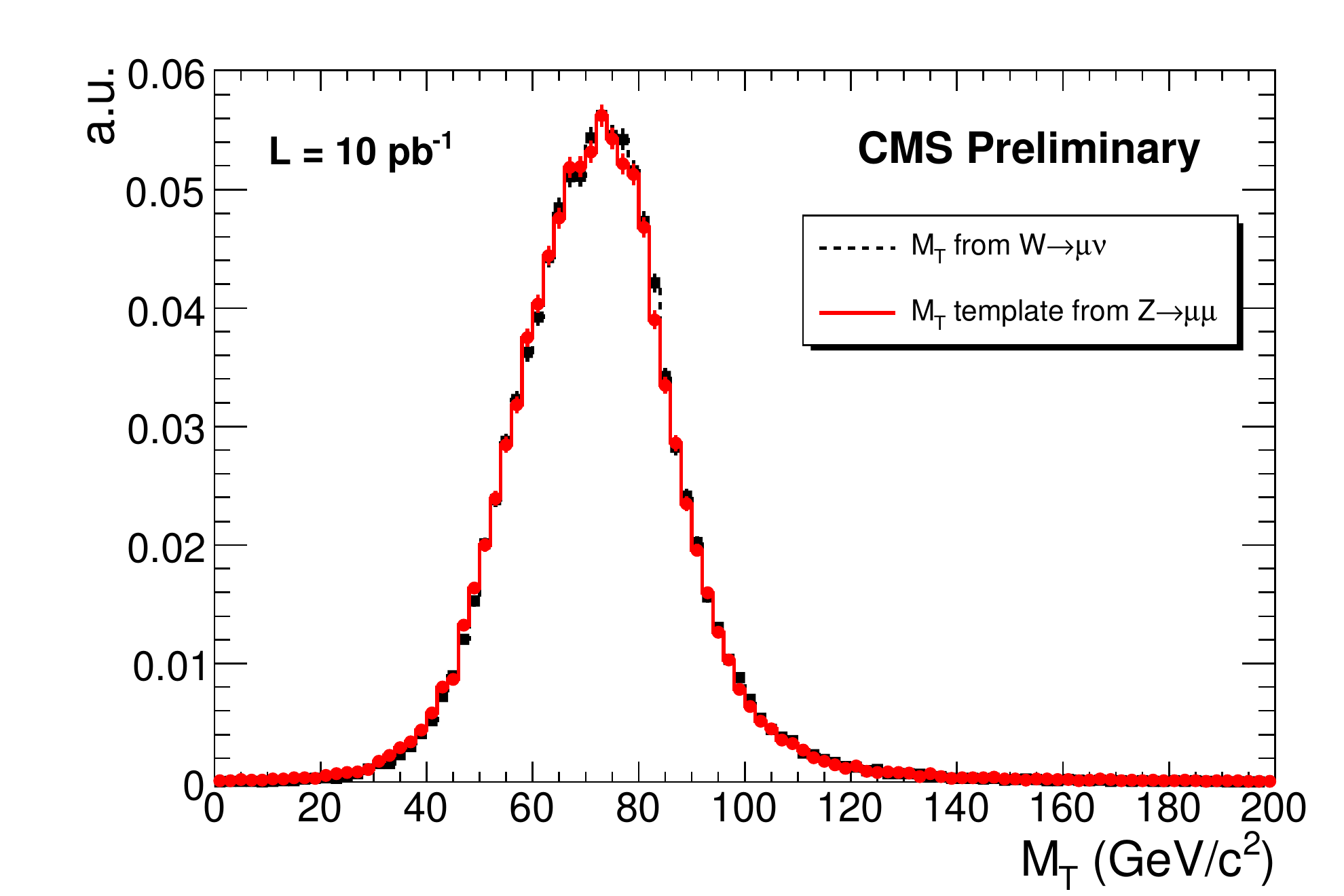} }
\subfigure[QCD background sample, produced by reverting the isolation cut ]{ \includegraphics[width=45mm,height=40mm]{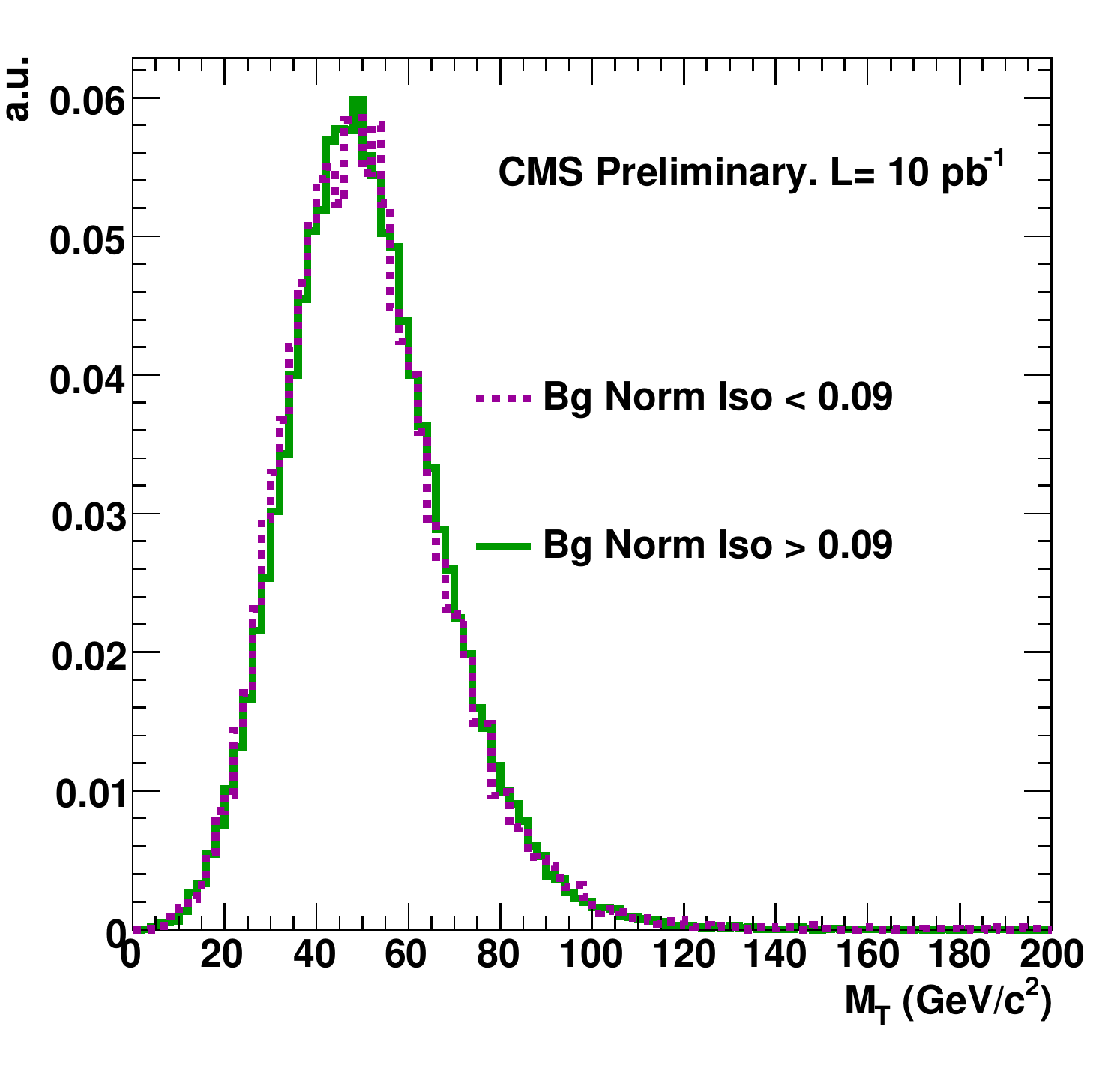} }
\end{center}
\end{figure}

Two variables can be used to fit the signal and background level: the lepton transverse momentum, and the lepton-neutrino transverse mass, shown in Figure~\ref{fig:transverse-mass}. The major issue for such analyses being the description of these distributions by the simulation, the CMS collaboration has shown how one can measure them using data. For the background, this is done by reverting the lepton isolation cut, to select a pure background control sample. For the signal, one uses $Z \rightarrow l^+ l^-$ events: if one of the leptons passes the $W \rightarrow l^+ \nu $ selection cut, the second lepton is ``removed'' from the event, thus creating fake missing energy. The transverse mass it then reconstructed. The mass difference between the Z and the W, as well as the neutrino acceptance are taken into account~\cite{CMS-notes}. Figure~\ref{fig:transverse-mass} shows the distributions obtained with 10 pb$^{-1}$. 

\section{Early physics studies using Z and W}

The reactions $pp \rightarrow W+X$ and $pp \rightarrow Z+X$ with subsequent W and Z leptonic decays will be among the first to be measured at the LHC. They are theoretically well understood: the calculation of the total cross-section at the NNLO has been done and the uncertainty due to the missing higher orders in perturbation theory is estimated to be less than 1~$\%$. Nevertheless, the uncertainty related to the use of different PDF parameterizations is estimated to be about 6~$\%$~\cite{TDR-CMS}. The uncertainty on the signal acceptance is evaluated to be about 2.3~$\%$~\cite{CSC-ATLAS,TDR-CMS}. The contribution to this number from the PDF uncertainty has been estimated by varying the eigenvectors of the same PDF parameterization (CTEQ6.5) and amounts to about 1~$\%$ ~\cite{CSC-ATLAS}.

ATLAS has estimated that the Z and W cross sections could be measured with a precision of 3 to 5~$\%$ with an integrated luminosity of 50 pb$^{-1}$. To this number, one must add the uncertainty on the luminosity, which is expected to be about 10~$\%$. With 1 fb$^{-1}$, the measurement precision should go down to 2.5~$\%$, and the uncertainty on the luminosity be 5~$\%$.

With 10 fb$^{-1}$, the precision on the W mass expected by CMS is $\delta m_W = 15 (stat) \oplus 30 (experimental) \oplus 10 (theoretical) = 35$ MeV. A recent study has revisited the various possible improvements and set a challenging goal of about 7 MeV per decay channel~\cite{prospects-W}. 
\noindent But the ATLAS collaboration has concentrated on the early measurements, with 15 pb$^{-1}$. In the electron channel, the fit of the transverse momentum distribution yields a precision of $ \delta m_W = 120 (statistical) \oplus 117 (systematical)$ MeV, where the systematics uncertainty is dominated by the electron energy scale. In the muon channel, the fit of the transverse mass gives $\delta m_W = 57 (statistical) \oplus 231 (systematical)$ MeV, where the dominant contribution comes from the calibration of the recoil energy . PDF uncertainties contribution is 25 MeV, other theoretical uncertainties are expected to be smaller. 

$W^+$ and $W^-$ are not produced with the same rapidity distribution. The $\eta$ distribution of the ratio $(W^+ - W^-)/(W^+ + W^-)$ is sensitive to PDF's. With 100 pb$^{-1}$, the measurement precision will be comparable to the present CTEQ6 predictions, and the first constraints will thus be set~\cite{CMS-notes}.

\section{Conclusion}

The ATLAS and CMS collaboration are obviously eagerly waiting for data, and have used the extra delay to make good progress on the realism of the detector simulation, the understanding of the expected performance, calibration and alignment studies, and to revisit the physics goals established over the years. Even with a reduced centre of mass energy of 10 TeV and about 200 pb$^{-1}$, the copious ``candle'' signals ( J/$\psi$, Z and W ) will allow us to shed light on the detector understanding, and perform some interesting measurements.


\section*{References}
\begin{thebibliography}{99}

\bibitem{commissionning} Masaya Ishino and Marina Giunta, same Conference proceedings.

\bibitem{ATLAS-paper} The ATLAS Experiment at the CERN Large Hadron Collider. JINST 3:S08003,2008.

\bibitem{CSC-ATLAS} Expected Performance of the ATLAS experiment, CERN-OPEN-2008-020. 

\bibitem{TDR-CMS} CMS Physics Performance TDR, CERN-LHCC-2006-021.

\bibitem{CMS-notes} CMS public notes BPH-070-002, EWK-07-002, EWK-08-002

\bibitem{prospects-W} N. Besson et al., Eur.Phys.J.C57:627-651,2008.

\end{thebibliography}
\end{document}